\documentclass[twocolumn,prb,showpacs]{revtex4}

\usepackage{epsfig}
\usepackage{graphics}
\usepackage{graphicx}
\usepackage{dcolumn}
\usepackage{bm}

\def\cm-1{cm$^{-1}$}

\def\sto{SrTiO$_{3}$\,}
\def\bbo{BaBiO$_{3}$\,}
\def\bkbo{Ba$_{1-x}$K$_{x}$BiO$_{3}$\,}

\begin{document}

\title
{
Surface Structure Analysis of Atomically Smooth \bbo Films
}

\author{A.~Gozar$^*$}
\author{G.~Logvenov}
\author{V. Y.~Butko}
\author{I.~Bozovic}
\affiliation{
Brookhaven National Laboratory, Upton, New York 11973-5000, USA
}


\begin{abstract}

Using low energy Time-of-Flight Scattering and Recoil Spectroscopy (TOF-SARS) and Mass Spectroscopy of Recoiled Ions (MSRI) we analyze the surface structure of an atomically smooth \bbo film grown by molecular beam epitaxy.
We demonstrate high sensitivity of the TOF-SARS and MSRI spectra to slight changes in the orientation of the ion scattering plane with respect to the crystallographic axes.
The observed angle dependence allows us to clearly identify the termination  layer as BiO$_2$.
Our data also indicate that angle-resolved MSRI data can be used for high resolution studies of surface structure of complex oxide thin films.

\end{abstract}


\maketitle

\emph{Introduction -}
The plethora of remarkable electrical and magnetic properties of transition metal oxides made them both a focus of basic research and very appealing candidates for integration in electronic devices.
For the latter, it is important to reproducibly synthesize and characterize atomically perfect surfaces and interfaces.
However, this goal is difficult to attain in complex oxides because of the complicated phase diagrams and high sensitivity to growth conditions.\cite{ivan-ieee}
This explains the large disparity between what is known about the structure of the very top atomic layers in these materials compared to widely used semiconductor or metal surfaces.

In this work we provide a new route to circumvent these problems.
This is done by combining the power of atomic layer-by-layer molecular beam epitaxy (ALL-MBE)~\cite{ivan-prl}, enabling production of films with perfect surfaces, with the extreme surface sensitivity of the low energy Time-of-Flight Scattering and Recoil Spectroscopy (TOF-SARS)~\cite{rabalais-book,bykov00} and the Mass Spectroscopy of Recoiled Ions (MSRI)~\cite{schultz-jvst99} techniques.
Angle-Resolved (AR) TOF-SARS could determine in principle inter-atomic spacings with a resolution approaching 0.01~\AA, comparable to lateral values obtained in surface X-ray crystallography and even better in the direction perpendicular to the surface.\cite{bykov00}
This comparison is to be judged also from the perspective of having a table-top experimental setup as opposed to the requirement of very intense synchrotron light.
However, in order to reach such accuracy one needs samples with very smooth surfaces. 
For this reason TOF-SARS and MSRI data in oxides have been primarily used for monitoring surface composition rather than the structure.\cite{auciello-jap06}
Exceptions are very few materials like Al$_{2}$O$_{3}$ or SrTiO$_{3}$ which are commercially available and commonly used as single crystal substrates for film growth.\cite{rabalais-alo,rabalais-sto} 

\begin{figure}[t]
\centerline{
\epsfig{figure=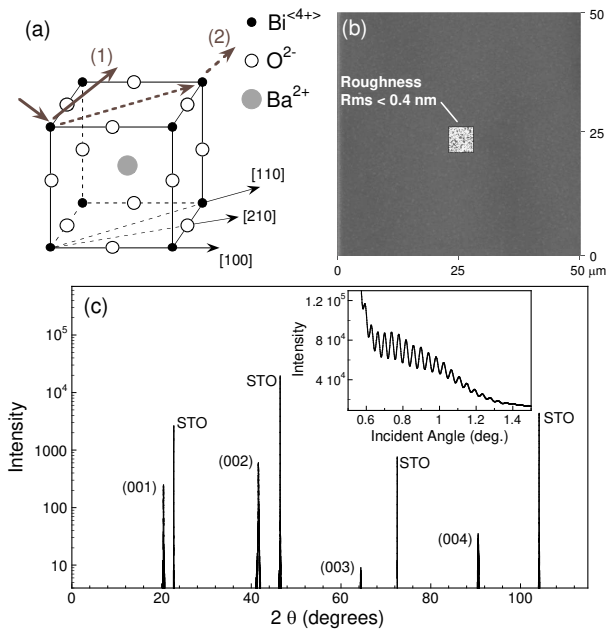,width=80mm}
}
\caption
{
(a) The cubic perovskite structure of undistorted \bbo.
(1) and (2) correspond to K trajectories as described in the text.
(b) A 50~$\times$~50~$\mu m$ AFM image of the \bbo film.
(c) $\omega - 2 \theta$ scan of the \bbo film; STO labels Bragg peaks from the \sto substrate.
The inset shows X-ray reflectance intensity oscillations at grazing incidence.
}
\label{f1}
\end{figure}

\bkbo is a family of superconductors with the maximum T$_c \simeq 32$~K (at x = 0.4) being the highest in an oxide material without copper.\cite{mattheiss,cava}
\bbo, the parent compound, is insulating and non-magnetic.
These interesting properties are believed to arise due to a charge-density-wave instability~\cite{pei} which leads to the lowering of the crystal symmetry from the simple cubic perovkite structure, see Fig.~\ref{f1}.
The driving force of this transition and the persistence of the charge order in superconducting \bkbo are still a matter of debate.\cite{pei,puchkov,thonhauser}
This is largely due to the difficulty in obtaining high quality single crystals from these materials.
A recent study brought to the forefront the important problem of dimensionality in \bkbo suggesting that this compound has in fact a layered structure, in analogy to the high T$_{c}$ superconducting cuprates.\cite{klinkova}
In the context of superconducting electronics, the interest in undoped and K doped \bbo stems from succesful fabrication of superconductor-insulator-superconductor tunnel junctions, a task that has bee quite ellusive with either cuprates or MgB$_2$ superconductors.
For these reasons it is critically important to understand and control the surface properties in this family of compounds.

\begin{figure}[t]
\centerline{
\epsfig{figure=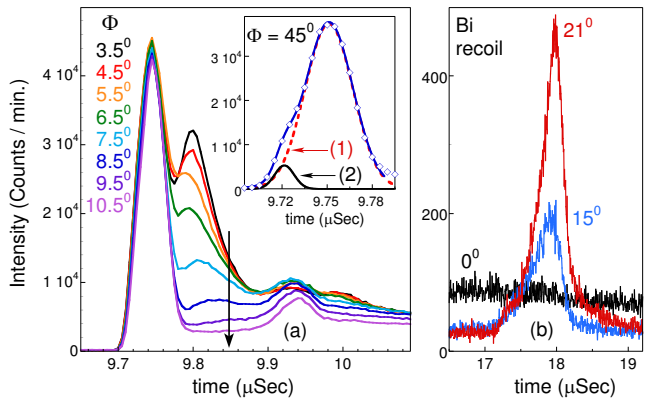,width=85mm}
}
\caption
{
(Color online)
Time-of-Flight K scattering data from \bbo recorded at $\theta = 27^{0}$ total scattering angle.
(a) The spectra are taken for azimuthal angles $\Phi$ from 3.5$^{0}$ to 10.5$^{0}$.
The arrow indicates the direction of increasing angle.
The peaks at $t = 9.74 \ \mu$s and $t \approx 9.8 \ \mu$s correspond to $K_S(Bi)$ and $K_S(Ba)$ respectively.
Inset: zoomed in area around the K$_{S}(Bi)$ peak for $\Phi = 45^{0}$.
The line through the data points is a two Gaussian peaks fit.
These two peaks, denoted by (1) and (2), correspond to the K trajectories shown in Fig.~\ref{f1}a.  
(b) The $Bi_{R}$ peak at higher times for $\Phi = 0^0$, $15^{0}$ and $21^{0}$.
}
\label{f2}
\end{figure}

\begin{figure}[t]
\centerline{
\epsfig{figure=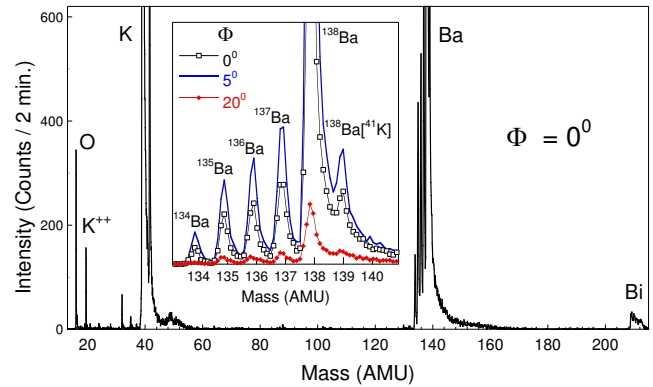,width=85mm}
}
\caption
{
(Color online)
The main panel displays a MSRI spectrum taken at $\Phi = 0^{0}$ azimuth.
The inset shows a zoomed in area of the Ba isotopes region from the main panel, for three values of $\Phi$: $0^{0}, \ 5^{0}$ and $20^{0}$.
}
\label{f3}
\end{figure}

Real space crystallography using TOF-SARS is based on the concepts of shadowing and blocking cones.\cite{rabalais-book,aono}
TOF-SARS is sensitive to both ions and neutrals so it is not dependent on charge exchange processes at the surface.
The drawback is that these spectra display broad 'tails' at high times (see Fig.~\ref{f2}) associated with multiple scattering events that are difficult to analyze quantitatively.
This drawback is eliminated by MSRI which achieves 'time focussing' according to $t = t_{0} + k (M / e)^{1/2}$, i.e. the flight time of the ions is only a  function of their mass to charge ratio.
The broad continua seen in TOF-SARS are turned into very sharp features allowing isotopic mass resolution~\cite{schultz-jvst99}, see Fig.~\ref{f3}.
MSRI is thus easier to interpret and one is tempted to use AR MSRI for surface structure analysis.
However, because MSRI detects only ions, one should worry about possible anisotropic neutralization effects which makes a \emph{quantitative} interpretation problematic.\cite{niehus} 
In fact this aspect is a long standing problem in ion based mass spectrometry affecting both compositional and structural studies.

Here we report on succesful and reproducible synthesis of large area single crystal thin films of \bbo using ALL-MBE.
This opens the path of improved basic experiments including high resolution surface crystallography in oxides based on ion scattering.
Next we present results of surface analysis of a \bbo film using AR TOF-SARS and MSRI.
We show that atomically smooth surfaces lead to high sensitivity of both type of spectra with respect to as small as one degree variation in the azimut angle.
Comparison between the AR TOF-SARS and MSRI data shows that the latter can be a powerful tool for quantitative surface structure analysis.
To the best of our knowledge this result has not been reported before.
The angular dependence of the spectra along with numerical simulations allows us to unambiguously determine that the \bbo film terminates by a BiO$_2$ layer.

\emph{Experimental -}
A \bbo thin film was grown on a \sto substrate.
The thickness of 96~nm was determined from  low angle X-ray reflectometry, a value in very good agreement with the prediction from the programmed growth rate.
Atomic force microscopy data show absence of any secondary phase precipitates and a roughness Rms that is below 4~\AA \ in a typical $25 \mu$m$^2$ area, see Fig.~1b.
This was expected based on our observation of strong specular reflection and undamped RHEED oscillations during growth.
The pronounced finite thickness reflectance oscillations (Fig.~\ref{f1}c - inset) clearly demonstrate atomically flat film interfaces.
The high quality of the film is furthermore reflected in the sharp Bragg diffraction peaks seen in the $\omega-2\theta$ scans shown in Fig.~\ref{f1}c.
The lattice constants of the bulk (96~nm thick) film at T~=~300~K were determined to be $c = 4.334(2)$~\AA \ while in-plane $a \approx b \simeq 4.353(4)$~\AA.
Note also that the unit cell of our \bbo film appears to be very close to the simple perovskite structure shown in Fig.~\ref{f1}b, which is not the case for the body-centered monoclinic structure characteristic of single crystals at this temperature.\cite{pei}
The experimental observation of excellent epitaxy is intriguing taking into account the relatively large mismatch, of about 11\%, between the in-plane lattice constants of \sto and bulk single crystals of \bbo.

The ion scattering data were taken using a recently built Ionwerks TOF system, based on principles described in Refs.\cite{schultz-jvst99,krauss-schultz}
We used a K$^{+}$ ion source tuned to provide a monochromatic beam of 10~keV.
Microchannel plate detectors were mounted at $\theta = 27^{0}$ and $37^{0}$ total scattering angles.
The incident angle was $\alpha = 15^{0}$.
Time resolved spectra were achieved by pulsing the source beam at a 20~kHz over a 0.5$\times$2~mm aperture.
The typical pulse width was 12~ns.
This, together with the $I \approx 0.1 \mu$A value of the ion current on the aperture, allows us to estimate an ion dose of about 3$\times$10$^{11}$ ions/cm$^{2}$ per spectrum.
The surface density is about 10$^{15}$ atoms/cm$^{2}$ indicating that the technique was not invasive for the duration of the experiment.
More important, measurements from pristine regions of the 1~cm$^2$ sample during and after the experiment ensured that the surface was neither damaged nor charged.
The TOF-SARS and MSRI spectra were collected at T~$\approx 600^{0}$~C in ozone atmosphere at a pressure $p \simeq$ 5$\times$10$^{-6}$~Torr.
In the following $K_{S}(X)$ denotes K$^{+}$ ions undergoing single scattering events from the element $X$ on the surface  while the notation $X_{R}$ stands for particles ($X = Ba, Bi$ or $O$) recoiled from the \bbo surface.
The numerical calculations were performed using the Scattering and Recoil Imaging Code (SARIC) which is a classical trajectory simulation program based on the binary collision approximation.\cite{saric}

\emph{Surface structure analysis of \bbo -}
Fig.~\ref{f2}a illustrates the dependence of the TOF-SARS spectra on the azimuth angle $\Phi$, defined as the angle between the scattering plane and the [100] direction of the cubic structure.
One minute long scans taken with $\Phi$ varied in 1$^{0}$ increments between 3.5$^{0}$ and 10.5$^{0}$ are shown.
The continuum above $t \ge 9.9 \ \mu$s is due to K multiple scattering.
Based on the predictions of elastic binary collision the two sharp features in Fig.~\ref{f2}a can be immediately assigned to single scattering events of K ions.
The peak at $t = 9.75$~ns corresponds to $K_{S}(Bi)$ and the one at $t \simeq 9.8$~ns to $K_{S}(Ba)$.
Dramatic changes are seen in the behavior of the higher time $K_{S}(Ba)$ peak as $\Phi$ is varied.

Surface roughness typically smears out the structure in AR scattering or recoil features. 
The strong sensitivity of the spectra in Fig.~\ref{f1} to small variations of the azimuthal angle is a consequence (and a direct proof) of high surface quality.
The angle independent intensity of the $K_{S}(Bi)$ peak along with the strong variation in the $K_{S}(Ba)$ feature suggest that the film terminates with BiO$_{2}$ planes which shadow the subsurface BaO layers.
We show below that this is fully supported by more detailed analysis including numerical simulations.

Information about the surface arrangement and dynamics can be obtained by studying the details of the spectral shapes.
For example the inset of Fig.~\ref{f2}a, which shows a zoomed in region around the $K_{S}(Bi)$ peak at $\Phi = 45^{0}$, reveals a shoulder on the low time side.
Indeed, the data can be well fitted by two Gaussian peaks which correspond to trajectories involving single and double scattering events denoted by (1) and (2) in Fig.~\ref{f1}a.
This assignment to events involving only K and Bi atoms along the [110] direction is based on the elastic binary collision model which predicts a difference of 33~ns between these two trajectories, in good agreement with the experimental value $\delta t = 30$~ns.
We do not observe the low time feature for $\Phi = 0^{0}$; this is understood as the effect of the intervening O atom along [100] direction.
Since the intensity of peak (2) depends strongly on the atomic arrangement at the surface, angular dependencies of the relative intensity of these two peaks could be used to get information about the symmetry and vibration amplitudes of the top layer atoms.
Shown in Fig.~\ref{f2}b is the $Bi_{R}$ peak which is found around $t \simeq 18 \mu$s.

We turn now to the question whether AR MSRI can be a quantitative probe for high resolution surface study.
It is natural to ask (a) if one can see structure in the AR MSRI data and (b) whether such dependence, if present, provides quantitative information about the surface.
The latter problem is related to the fact that it is hard to quantify the yield of ionic fraction which, moreover, could be itself an anisotropic function with respect to the orientation of the crystallographic axes.\cite{niehus}

The answer to the first question is given in Fig.~\ref{f3}.
The main panel shows a MSRI spectrum taken at $\Phi = 0^{0}$ and the inset shows the Ba isotopes region for three azimuths.
Clearly, the intensities of the corresponding peaks vary substantially when this angle is changed.
Note also the sharpness of the peaks which allows for easy separation of atomic isotopes.
The mass resolution, $m / \Delta m \approx 380$, is about one order of magnitude better than that obtained in typical TOF-SARS spectra.

\begin{figure}[t]
\centerline{
\epsfig{figure=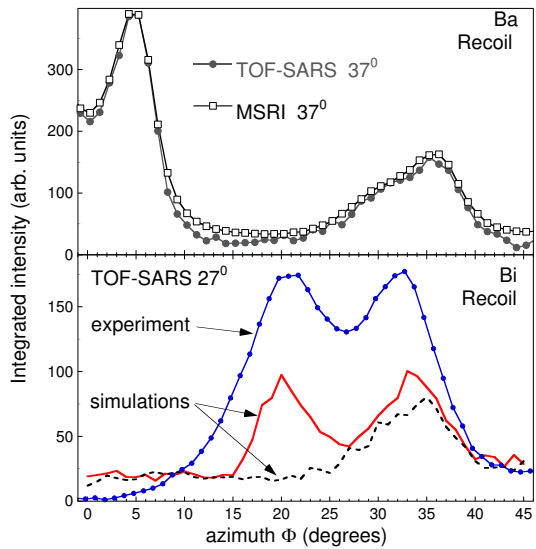,width=70mm}
}
\caption
{
(Color online) Top: azimuth dependence of the integrated intensity of the Ba recoil peak from two sets of measurements taken at $\theta = 37^0$ scattering angle: MSRI (squares) and TOF-SARS spectra (circles).
The curves are matched at $\Phi = 5^{0}$.
Bottom: $\Phi$ dependence of the experimental integrated intensity of the $Bi_{R}$ peak from TOF-SARS data of Fig.~\ref{f2}b.
The solid and dashed lines correspond to SARIC simulations assuming BiO and a BaO film terminations respectively.
The data in this panel are matched at $\Phi = 45^{0}$.
}
\label{f4}
\end{figure}

The second question is addressed in the top panel of Fig.~\ref{f4} where two data sets are compared.
One data set corresponds to the $\Phi$ dependence of the intensities of MSRI $Ba_R$ peaks derived from the spectra shown in the inset of Fig.~\ref{f3}.
The second data set refers to  $Ba_R$ peak from the TOF-SARS spectra taken with the detector at the same total scattering angle $\theta = 37^{0}$.
The latter data set was taken with the time focussing analyser not biased, making it sensitive to both ion and neutral particles.
The two dependencies are essentially identical.
Equally good agreement is also observed if the $Bi_{R}$ feature is considered instead of $Ba_{R}$.
We conclude that anisotropic neutralization effects are not important which proves that AR MSRI can be used as a quantitative probe for surface analysis.
The insulating nature of the film and the use of alkali ion source beam could be responsible for this effect.\cite{niehus}
Note also that the steep decrease in the $Ba_{R}$ signal between $\Phi = 5^{0}$ and $10^{0}$ seen in Fig.~\ref{f3} is consistent with the disappearance of the $K_S(Ba)$ peak at $t \simeq 9.8 \mu$s in the TOF-SARS data at $\theta = 27^{0}$ from Fig.~\ref{f2}a.

We address now the problem of film surface termination.
The spectra shown in the bottom panel of Fig.~\ref{f4} provide the answer to this question.
The experimental points correspond to the $\Phi$ dependence of the integrated intensity of the $Bi_{R}$ peak shown in Fig.~\ref{f2}b.
For every data point we subtracted the background due to the high time tail associated with K multiple scattering (as seen in Fig.~\ref{f2}a).
The results of two SARIC simulations are also shown in Fig.~\ref{f4}.
The solid and dashed lines correspond to assumed BiO$_{2}$ and BaO terminations respectively.
The simulation based on BiO$_{2}$ termination reproduces well the main features of the experiment, i.e. the two peaks around 20$^{0}$ and 33$^{0}$.
In contrast, the lower angle feature is absent if BaO is assumed to be the topmost layer.
These numerical simulations clearly show that the film terminates in a BiO$_{2}$ surface.  

One advantage of real space structure analysis is the intuitive picture it immediately provides: the dips in the azimuth scans generally correspond to low index crystallographic directions in the material.
In Fig.~\ref{f4}b the minima in the experimental scan found at $\Phi = 0^{0}, 27^{0}$ and $45^{0}$ are associated with the [100], [210] and [110] directions of the cubic structure respectively, see Fig.~\ref{f1}.

Because these effects strongly depend on the interatomic distances as well as the type of atoms on the surface, the magnitude of these minima can also be explained qualitatively.
The severe shadowing and bloking due to both Bi and O atoms lying along the [100] azimuth are the cause of the absence of $Bi_{R}$ signal for $\Phi = 0^{0}$.
Although Oxygen atoms contribute to this effect, due to their lower mass they cannot completely shadow or block the incoming K ions or the recoiled Bi along [210] direction.
As a result only a shallow minimum is seen around $\Phi = 27^{0}$.
This is not the case for Bi atoms which are responsible for the more pronounced dip at $\Phi = 45^{0}$ in spite of the larger interatomic separation along the [110] direction.
Similarly, the absence of a peak for $\Phi \simeq 20^{0}$ in the simulated $Bi_{R}$ intensity for assumed BaO termination (dashed line in Fig.~\ref{f4}b) can be easily understood in terms of Ba shadowing effects on Bi atoms along the [141] direction and the value $\alpha = 15^{0}$ for the incidence angle.
A detailed analysis of the lattice constants and possible surface relaxation based on numerical calculations of the shadowing and blocking cones are the purpose of future work.


\emph{Conclusions -}
Achieving atomically smooth surfaces is proven to have a great impact on the possibility to use AR TOF-SARS and MSRI for structure and interface analysis in complex oxides.
This is a stepping stone for future characterization of artificially layered superconducting compounds like \bkbo or the cuprates.
The quantitative agreement between TOF-SARS and MSRI spectra of a \bbo film show that AR MSRI can be a powerful tool for high resolution surface analysis. 
Data and simulations allowed us to clearly identify that the \bbo film terminates in a BiO$_2$ layer.

\emph{Acknowledgements -}
This work was supported by DOE grant DE-AC-02-98CH1886.
We thank W.~J. Rabalais for providing us with the SARIC simulation code and J.~A. Schultz for useful discussions.

\end{document}